\documentclass[12pt]{iopart}

\usepackage{iopams}
\usepackage{graphicx}
\def\a{true} \ifx\a\useyap
\usepackage[dvipdf,
           colorlinks=true,
            citecolor=blue,
            linkcolor=blue,
          anchorcolor=black,
             urlcolor=black
           ]{hyperref}
\else
\usepackage[dvipdfm,
            colorlinks=true,
             citecolor=blue,
             linkcolor=blue,
           anchorcolor=blak,
              urlcolor=black
            ]{hyperref}

\begin{document}

\title
{The finite-width Laplace sum rules for $0^{++}$ scalar glueball in instanton liquid model}

\author{Shuiguo Wen, Zhenyu Zhang and Jueping Liu\footnote[7]{Corresponding author}}

\address{College of Physics and Technology, Wuhan University, 430072 Wuhan,
China}
\eads{\mailto{airywsg@163.com}, \mailto{zyzhang@me.com} and \mailto{jpliu@whu.edu.cn}}

\begin{abstract}
In the framework of a semi-classical expansion for quantum
chromodynamics in the instanton liquid background, the correlation
function of the $0^{++}$ scalar glueball current is given. Besides
the pure classical and quantum contributions, the contributions
arising from the interactions between the classical instanton fields
and quantum gluons are taken into account as well. Instead of the
usual zero-width approximation for the resonance, the Brite-Wigner
form for the spectral function of the finite-width resonance is
adopted. The family of the Laplace sum rules for the scalar glueball
in quantum chromodynamics with and without light quarks are studed.
A consistency between the subtracted and unsubtracted sum rules are
very well justified, and the values of the mass, decay width, and
the coupling to the corresponding current for the $0^{++}$ resonance
in which the glueball fraction is dominant, are obtained.
\end{abstract}

\maketitle

\section{Introduction}
\label{sec:introduction}

Understanding the nature of the lightest state of glueballs, the
$0^{++}$ scalar glueball, is a long-standing puzzle in
QCD\cite{KZ07,MKV09}. The mass scale of this glueball is
predicted to be within the region of 1.30 -1.75GeV by
quenched Lattice QCD\cite{Colin99,Meng09,Chen06,Vaccar99,Sexton95}, and by un-
quenched lattice QCD\cite{SESAM00,McNeile01,Hart02}. Up to now, there has not been clear evidence for the
observation of a scalar glueball, while the closest scalar
candidates are $f_0(1370)$, $f_0(1500)$ and $f_0(1700)$\cite{KZ07,
CM09}. More nonperturbative physics is needed in the theoretical
and phenomenological investigation on such area.

Laplace sum rule\cite{SVZ79} calculations of glueball properties can
be based on correlation functions involving interpolating field,
which is not as much successful in the prediction of scalar glueball
mass as in the ones of other hadron properties in the early days
with the inconsistency between the  subtracted\cite{NSVZ80, JPLiu93,
JPLiu91, Badesd89, Doming86} and unsubtracted\cite{Shifman81,
Pascual82} sum rules(SSR and USSRs). Instanton vacuum should be
included in the scalar tunnel\cite{NSVZ81, GI80, Shuryak83} to offer
the main non-perturbative effects.

Instantons are the localized solutions of the classical Euclidean
field equations with finite minimized action \cite{BPST75}. They can
be solved by constructing the self-dual or self-antidual field
configurations classified by different topological charges. The
perturbative theory should be carried out around these classical
solutions with the average zero-topological charge instead of the
trivial one, which is the kernel of the semiclassical expansion
method.

Not all the hadrons alike\cite{NSVZ81}. Instanton contributions
should not be neglected at least in the scalar and pseudoscalar
channels. Direct instanton contributions are already included in sum
rule approaches\cite{Forkel05, Forkel01, Sch95, KJ01, HSE01, HaS01}
based on the instanton liquid model of the QCD vacuum\cite{Dia03,
SS98}. The compatibility between the resultant USSRs and SSR for the
scarlar glueball is greatly improved, but still not be very
satisfying. It should be noticed that, in the so-called direct
instanton approximation, the interactions between instantons and the
pure quantum gluon fields are not considered, and the procedure is
criticized by involving with the problem of double
counting\cite{Forkel01}, because in the correlator are included both
contributions coming from condensates and instantons, but the latter
could lead to the formation of the former.

The interactions between the classical and quantum gluon field
configurations are always ignored\cite{Forkel05, Forkel01, Sch95,
KJ01, HSE01, HaS01} because the interactive effects were expected to
be small\cite{Sch95}. However, there is no reasonable argument
before an actual calculation. We have found that these interactive
effects are, in fact, compatible with or even important more than
that of the condensate and the perturbative effects at least in the
$0^{++}$ channel. Moreover, including the classical, quantum and
interactive effects in the framework of the semiclassical expansion
of the instanton background, the stability and the consistency for
the SSR and USSRs for $0^{++}$ scalar glueball could be arrived
\cite{ZhangZY06}.

Motivated by the above considerations, our main purpose in this
paper is to investigate the $0^{++}$ glueball in the frame work of
Laplace sum rules. To avoid the problem of double counting, instead
of using the scheme of the mixture of the traditional condensates
and the so-called direct instanton contribution, we are working in
the framework of the semiclassical expansion of QCD in the instanton
liquid vacuum, which is a well-defined self-consistent procedure for
the quantum theory justified by the path-integral quantization
formalism. For the correlation function, we include the
contributions from the interactions between the quantum gluons and
the classical instanton background besides the ones coming from only
instantons and from only quantum gluons. For the spectral function,
beyond the usual zero-width approximation, we adopt the Breit-Wigner
form for the considered resonance with correct threshold behavior,
in order to get the information of not only the mass scale but also
the full decay width.

\section{Correlation function}
\label{chap:1} The correlation function for the scalar glueball in
Euclidean space-time with the virtuality $q^2$ is defined by
\begin{equation}
\Pi(q^2)=\int\textrm{d}^4x\textrm{e}^{iq\cdot
x}\langle\Omega|TO_s(x)O_s(0)|\Omega\rangle\label{eq:COR},
\end{equation}
where $|\Omega\rangle$ is the physical vacuum, and the scalar
glueball current $O_s$ of the quantum numbers $J^{PC}=0^{++}$ is
given by
\begin{equation}
O_s=\alpha_sG^a_{\mu\nu}(B)G^{a,\mu\nu}(B)\label{eq:current},
\end{equation}
with $\alpha_s$ being the strong coupling constant, it is
gauge-invariant, and renormalization invariant at one-loop-level. On
spirit of the semiclassical expansion, and in order to maintain the
$O(4)$-covariance, the gluon field strength tensor $G^a_{\mu\nu}(B)$
is considered as a functional of the full gluon potential $B_{\mu
a}=A_{\mu a}+a_{\mu a}$ with $A_{\mu a}$ and $a_{\mu a}$ being the
instanton fields and the corresponding quantum fluctuations.

The theoretical expression, $\Pi^{\textrm{QCD}}$, for the
correlation function $\Pi$ may be divided into the following three
parts
\begin{equation}
\Pi^{\textrm{QCD}}(Q^2)=\Pi^{\textrm{inst}}(Q^2)+\Pi^{\textrm{int}}(Q^2)+\Pi^{\textrm{pert}}(Q^2),
  \label{eq:PQCD}
\end{equation}
where $Q^2=q^2$, and $\Pi^{\textrm{inst}}(Q^2)$,
 $\Pi^{\textrm{pert}}(Q^2)$, and $\Pi^{\textrm{int}}(Q^2)$ are the contributions from the pure instatons,
the pure perturbation QCD, and the interactions between the
instantons and the quantum gluon fields, respectively. We note that
we have not included here the contributions from the so-called
condensates because at first in a systematic semiclassical expansion
of QCD, the non-perturbative effects are parameterized by the
classical instanton and anti-instanton solutions of the equation of
motion of QCD, and at second we want to avoid the double counting
problem due to the fact that some condensates can be reproduced from
the instanton contributions, and thirdly, we have checked that the
condensates contributions are negligible in comparison with the
contributions considered here.

The perturbative contribution $\Pi^{\textrm{pert}}(Q^2)$ up to
three-loop level in the chiral limit of QCD is already known to be
\begin{equation}
\Pi^{\textrm{pert}}(Q^2)=Q^4\ln\left(\frac{Q^2}{\mu^2}\right)
\left[a_0+a_1\ln\left(\frac{Q^2}{\mu^2}\right)
+a_2\ln^2\left(\frac{Q^2}{\mu^2}\right)\right]\label{eq:pert},
\end{equation}
where $\mu^2$ is the renormalization scale in the
$\overline{\mbox{MS}}$ dimensional regularization scheme, and the
coefficients with the inclusion of the threshold effects are
\begin{eqnarray}
a_0&=&-2\left(\frac{\alpha_s}{\pi}\right)^2\left[1+\frac{659}{36}
\left(\frac{\alpha_s}{\pi}\right)+247.48\left(\frac{\alpha_s}{\pi}\right)^2\right]
,\nonumber\\
a_1&=&2\left(\frac{\alpha_s}{\pi}\right)^3\left[\frac9{4}+65.781\left(\frac{\alpha_s}{\pi}\right)\right],\quad
a_2=-10.1252\left(\frac{\alpha_s}{\pi}\right)^4\label{eq:pert3}
\end{eqnarray}
for QCD with three quark flavors up to three-loop level in the
chiral limit \cite{HSE01,HaS01,CKS97}, and
\begin{eqnarray}
a_0=-2\left(\frac{\alpha_s}{\pi}\right)^2\left[1+\frac{51}{4}
\left(\frac{\alpha_s}{\pi}\right)\right],\quad
a_1=\frac{11}{2}\left(\frac{\alpha_s}{\pi}\right)^3,\quad
a_2=0 \label{eq:pert2}
\end{eqnarray}
for quarkless QCD up to two-loop level \cite{Bagan90}. Both
expressions for $\Pi^{\textrm{pert}}(Q^2)$ with and without quark
loop corrections are used in our calculation for comparison. With
the assumption that the dominant contribution to
$\Pi^{\textrm{inst}}(Q^2)$ comes from BPST single instanton and
anti-instanton solutions \cite{BPST75, tHft76,CDG78} and the
multi-instanton effects are negligible \cite{Sch95}, and in view of
the gauge-invariance of the correlation function, one may choose to
work in the regular gauge of the classical single instanton
potential
\begin{eqnarray}
 A^a_{\mu}=\frac{2}{g_s}\eta^a_{\mu \nu}\frac{(x-x_0)_{\nu}}{(x-x_0)^2+\rho^2},
\end{eqnarray}
where $\eta^a_{\mu \nu}$ is the 't Hooft symbol, and $x_0$ and
$\rho$ denote the position and size of the instanton, respectively.
The pure instaton contribution is obtained to be
\cite{NSVZ80,Forkel05,Forkel01,HSE05,Shu82,Ges80,Ioffe00}
\begin{equation}
\Pi^{\textrm{inst}}(Q^2)=2^5\pi^2\bar{n}\bar{\rho}^4Q^4K^2_2(\sqrt{Q^2}\bar{\rho})\label{eq:inst},
\end{equation}
where $K_2(x)$ is the McDonald function, $\bar{n}=\int_0^\infty
d\rho n(\rho)$ and $\bar{\rho}$ are the overall instanton density
and the average instanton size in the random instanton background,
respectively.

It is noticed that the contribution to $\Pi$ from the interactions
between instantons and the quantum gluon fields is of the order of
the product of $\alpha_s$ and the overall instanton density
$\bar{n}$. There is no reason to get rid off this contribution in
comparison with the perturbative contributions of the higher order
$\alpha_s^4$ considered in $\Pi^{\textrm{pert}}$. To calculate such
contribution, our key observation is that the instanton potential
$A^a_{\mu}$ obeys also the fixed-point gauge condition
\begin{equation}
 (x-x_0)_{\mu}A^a_{\mu}(x-x_0)=0,
\end{equation}
due to the anti-symmetricity of the 't Hooft symbols. As a
consequence, the instanton potential can be expressed in terms of
the corresponding field strength tensor as follows
\begin{equation}
 A^a_{\mu}(x-x_0)=\int_0^1duuF^a_{\mu \nu}[u(x-x_0)](x-x_0)_{\nu},
   \label{eq:gauge-condition}
\end{equation}
and the gauge-link with respect to the instanton fields is just the
unit operator, and thus the trace of any product of the
gauge-covariant instanton field strengths at different points is
gauge-invariant. This allows us to conclude that the remainder
quantum corrections to the gauge-invariant correlation function,
arising from the interactions between the instantons and the quantum
gluons, is gauge-invariant as well. Therefore, one may choose any
specific gauge in evaluating the quantum correction to
$\Pi^{\textrm{int}}$. Working in Feynman gauge, our results for
$\Pi^{\textrm{int}}$ is
\begin{eqnarray}
 \Pi^{\textrm{int}}(Q^2)=C_0\alpha_s\bar{n}\pi+\alpha_s^2\bar{n}\left\{C_1
 +[C_2(Q\bar{\rho})^2+C_3]\ln(Q\bar{\rho})^2+\frac{C_4}{(Q\bar{\rho})^{2}}\right\}, \label{eq:int}
\end{eqnarray}
where the coefficients are:
\begin{eqnarray}
 &&C_0=62.62,\,\,  C_1=1533.15,\,\, C_2=825.81,\nonumber\\
 &&C_3=-496.33,\,\, C_4=-348.89.
\end{eqnarray}
It is remarkable to note that the fixed point $x_0$, which
characterizes the gauge condition (\ref{eq:gauge-condition}),
disappears in the expression of $\Pi^{\textrm{int}}$, as expected
from the gauge-invariance of our procedure.

\section{Spectral function}
\label{chap:2} Turn to construct the spectral function for the
correlation function of the scalar glueball current. The usual
lowest one resonance plus a continuum model is used to saturate the
phenomenological spectral function,
\begin{equation}
\textrm{Im}\Pi^{\textrm{PHE}}(s)=\rho^{\textrm{had}}(s)
+\theta(s-s_0)\textrm{Im}\Pi^{\textrm{QCD}}(s)
  \label{eq:ImPHE},
\end{equation}
where $s_0$ is the QCD-hadron duality threshold,
$\rho^{\textrm{had}}(s)$ the spectral function for the lowest scalar
glueball state, and the imaginary part of the correlation function
Eq. (\ref{eq:PQCD}), $\textrm{Im}\Pi^{\textrm{QCD}}(s)$, is
\begin{eqnarray}
\textrm{Im}\Pi^{\textrm{QCD}}(s)&=&-\pi
s^2\left[a_0+2a_1\ln \frac{s}{\mu^2}+\left(3\ln^2 \frac{s}{\mu^2}-\pi^2\right)a_2\right]\nonumber\\
&&-16\pi^4s^2\bar{n}\bar{\rho}^4J_2(\bar{\rho}\sqrt{s})Y_2(\bar{\rho}\sqrt{s})
+\alpha_s^2\bar{n}\pi(C_2\bar{\rho}^2s-C_3) \label{eq:ImQCD}.
\end{eqnarray}
Instead of using the zero-width approximation as usual, the
Brite-Wigner form for $n$ resonances is adopted for
$\rho^{\textrm{had}}(s)$
\begin{equation}
\rho^{\textrm{had}}(s)=\sum_{i=1}^{n}\frac{f_i^6m_i\Gamma_{i}}{(s-m^2_{i}
+\Gamma^2_{i}/4)^2+m_{i}^2\Gamma_{i}^2}
  \label{eq:had2},
\end{equation}
where $f^3_i=\langle\Omega|O_s|0^{++}_i\rangle$ is the coupling of
the $i$'s resonance to the glueball current (\ref{eq:current}).
Recall the threshold behavior for $\rho^{\textrm{had}}(s)$
\begin{equation}
 f^3_i\rightarrow \lambda_0s \hspace{20pt}\mbox{for } s\rightarrow 0,
   \label{eq:threshold}
\end{equation}
with the value of $\lambda_0$ being fixed by the low-energy theorem
of QCD\cite{NSVZ80,Shifman81,NSVZ81,Novikov81}, and thus independent
of what an individual resonance considered. The early QCD sum rule
approach had often used $f^3_i\rightarrow \lambda_0s$ (with $n=1$)
in the whole lowest resonance region, however the obtained mass
scale is too low to be expected from lattice QCD simulations. In
fact, the threshold behavior (\ref{eq:threshold}) is valid in the
chiral limit, it may not be extrapolated far away. Therefore,
instead of considering the couplings $f_i$ as constants
\cite{Sch95}, we choose the model for $f$ as
\begin{equation}
 f^3_i=
 \left\{
 \begin{array}{ll}
  \lambda_0s        &\mbox{for } s<m^2_{\pi},\\
  \lambda_0m^2_{\pi}+\lambda^3_i &\mbox{for } s\geq m^2_{\pi},
 \end{array}\right.  \label{eq:lambda}
\end{equation}
with $\lambda_i$ being some constants, so that the spectral function
$\rho^{\textrm{had}}(s)$ has the almost complete Breit-Wigner form
with correct threshold behavior which is important to maintain the
convergence of the integral for the spectral function of the $k=-1$
sum rule.

\section{Finite-width Laplace sum rules}
\label{chap:3} A family of Laplace sum rules with different
$k$-moments can be constructed from the Borel transformation,
$\hat{\mathcal{B}}$, to the correlation function
(\ref{eq:PQCD})\cite{SVZ79}
\begin{equation}
\mathcal{L}^{\textrm{had}}_{k}(s_0,t)
=\mathcal{L}^{\textrm{QCD}}_{k}(s_0,t)+\Pi(0)\delta_{k,-1},
   \label{eq:LSR}
\end{equation}
where $s_0$ is the threshold for setting on the continuum, $\Pi(0)$
is come from the subtraction to the corresponding dispersion
relation due to the degree of divergence of the correlation function
of the scalar glueball, and
\begin{eqnarray}
 \mathcal{L}^{\textrm{QCD}}_k(s_0,t)&=& t\hat{\mathcal{B}}
 \left[(-Q^2)^k\Pi^{\textrm{QCD}}(Q^2)\right]-\int^{\infty}_{s_0}\textrm{d}ss^k
 e^{-s/t}\frac1{\pi}\Pi^{\textrm{QCD}}(s),\\
 \mathcal{L}^{\textrm{had}}_{k}(s_0,t) &=&\int^{s_0}_0\textrm{d}ss^k
 e^{-s/t}\frac1{\pi}\rho^{\textrm{had}}(s).
\end{eqnarray}
The Laplace sum rule emphasizes the contribution from the lowest
hadron state considered, and suppresses the higher resonance
contributions and the continuum exponentially.

For $k=-1,0$ and $1$, a straightforward manipulation leads to
\begin{eqnarray}
\mathcal{L}^{\textrm{QCD}}_{-1}(t)&=&
\left[-a_0+(2\gamma-2)a_1-0.5(6\gamma^2-12\gamma-\pi^2)a_2\right]t^2+2^7\pi^2\bar{n}\nonumber\\
&&-2^6\pi^2\bar{n}x^2\textrm{e}^{-x}\left[(1+x)K_0(x)
+\left(2+x+\frac2{x}\right)K_1(x)\right]\nonumber\\
&&-C_0\bar{n}\pi\alpha_s
+\bar{n}\alpha_s^2[-C_1+C_2\bar{\rho}^2t+C_3(\gamma-\ln(\bar{\rho}^2t))-\frac{C_4}{\bar{\rho}^2t}],
\end{eqnarray}
\begin{eqnarray}
\mathcal{L}^{\textrm{QCD}}_{0}(t)&=&
\left[-2a_0+(4\gamma-6)a_1-(6\gamma^2-18\gamma-\pi^2+6)a_2\right]t^3\nonumber\\
&&+2^7\pi^2\bar{n}\frac{x^5}{\bar{\rho}^2}\textrm{e}^{-x}\left[2K_0(x)
+\left(2+\frac1{x}\right)K_1(x)\right]\nonumber\\
&&+\bar{n}\alpha_s^2[C_2\bar{\rho}^2t^2-C_3t+C_4/(\bar{\rho}^2)],
\end{eqnarray}
\begin{eqnarray}
\mathcal{L}^{\textrm{QCD}}_{1}(t)&=&
\left[-6a_0+(12\gamma-22)a_1-(18\gamma^2-66\gamma-3\pi^2+36)a_2\right] t^4\nonumber\\
&&+2^8\pi^2\bar{n}\frac{x^6}{\bar{\rho}^4}\textrm{e}^{-x}\left[(9-4x)K_0(x)
+\left(7-4x+\frac3{x}\right)K_1(x)\right]\nonumber\\
&&+\bar{n}\alpha_s^2[2C_2\bar{\rho}^2t^3-C_3t^2],
\end{eqnarray}
where $x=\bar{\rho}^2t/2$.

\section{Numerical calculation}
Now, we specify the input parameters in numerical calculation. We
take the color and flavor numbers to be $N_c=3$ and $N_f=3$,
respectively. The expressions for two-loops quarkless ($N_f=0$)
running coupling constant $\alpha_s(Q^2)$ at renormalization scale
$\mu$ \cite{Groom00,Prosperi06} and for the three-loop running
coupling constant with three flavors ($N_f=3$) are used, where the
central value of the $\overline{\textrm{MS}}$ QCD scale $\Lambda$ is
taken to be $120$ MeV. We recall here that a research on the
renormalization group improvement for Laplace sum rules amount to
choose the renormalization scale to be $\mu^2=t$ \cite{Narison81}.
The subtraction constant $\Pi(0)$ is determined by low-energy
theorem \cite{NSVZ80}
\begin{equation}
\Pi(0)=\frac{32\pi}{9}\langle\alpha_s
G^2\rangle\simeq0.6\,\textrm{GeV}^4,
\end{equation}
which leads to $\lambda_0=5.0$GeV. The values of the average
instanton size and the overall instanton density are adopted from
the instanton liquid model
\begin{equation}
\bar{n}=1\,\textrm{fm}^{-4}=0.0016\,\textrm{GeV}^4,\,\,\,
\bar{\rho}=\frac1{3}\,\textrm{fm}=1.689\,\textrm{GeV}^{-1}.
\end{equation}
Finally, the mass of the neutral pion is taking from the
experimental data, i.e. $m_{\pi}=135$ MeV.

To determining the values of the resonance parameters appearing in
Eq.(\ref{eq:had2}), we match both sides of sum rules (\ref{eq:LSR})
optimally in the fiducial domain. The conditions for determining the
value of $s_0$ are: first, it should be grater than $m^2$; second,
it should guarantee that there exists a sum rule window for our
Laplace sum rules. We note that the upper limit $t_{\textrm{max}}$
of the sum rule window is determined by requiring that the
contribution from the continuum should be less than that of the
resonance 
\begin{equation}
\mathcal{L}^{\textrm{cont}}_{k}(s_0,t_{\textrm{max}})\leq
\mathcal{L}^{\textrm{QCD}}_{k}(s_0,t_{\textrm{max}}), 
\end{equation}
while the lower limit $t_{\textrm{min}}$ of the sum rule window is obtained by
requiring the contribution of pure instantons to be greater than
50\% of $\mathcal{L}^{\textrm{QCD}}_k(s_0,t)$, because such
classical contributions should be dominant in the low-energy region.
Moreover, to require that the multi-instanton corrections remain
negligible, we simply adopt a rough estimate
\begin{equation}
t_{\textrm{min}}^{-1}\leq (2\bar{\rho})^2\sim
 ({2}/{0.6\textrm{GeV}})^2.
\end{equation}
In order to measure the compatibility between both sides of the sum
rules (\ref{eq:LSR}) realized in our numerical simulation, we
introduce a variation, $\delta$, defined by
\begin{equation}
\delta=\frac1{N}\sum_1^N\frac{[L(t_i)-R(t_i)]^2}{|L(t_i)R(t_i)|},
\end{equation}
where the interval $[t_{\textrm{min}}, t_\textrm{max}]$ is divided into
$100$ equal small intervals, $N=101$, and $L(t_i)$ and $R(t_i)$ are
l.h.s and r.h.s of Eq.(\ref{eq:LSR}) evaluated at $t_i$.

\begin{figure}[!hbp]
\begin{center}
\includegraphics*[angle=0,width=8.5cm]{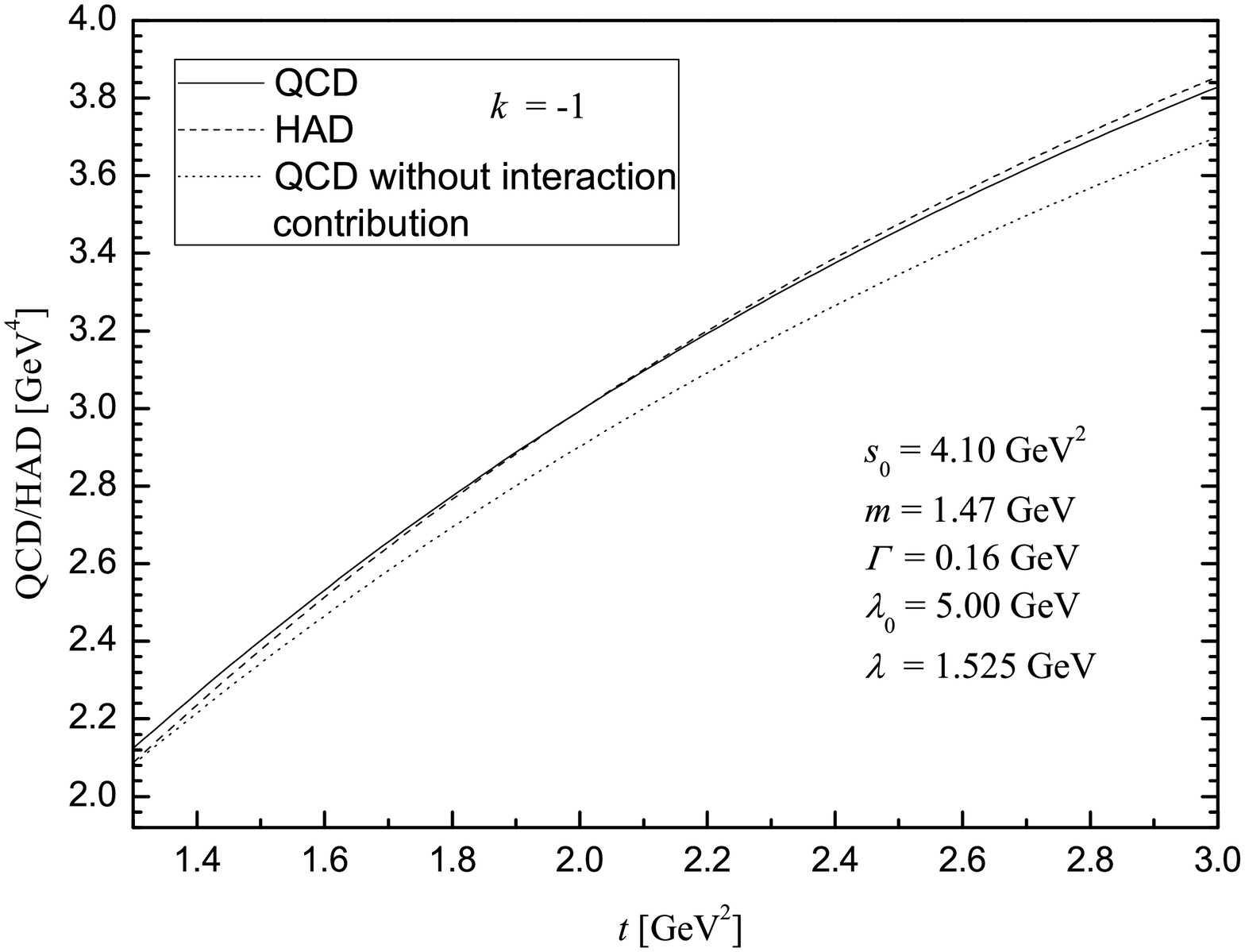}\\
\includegraphics*[angle=0,width=8.5cm]{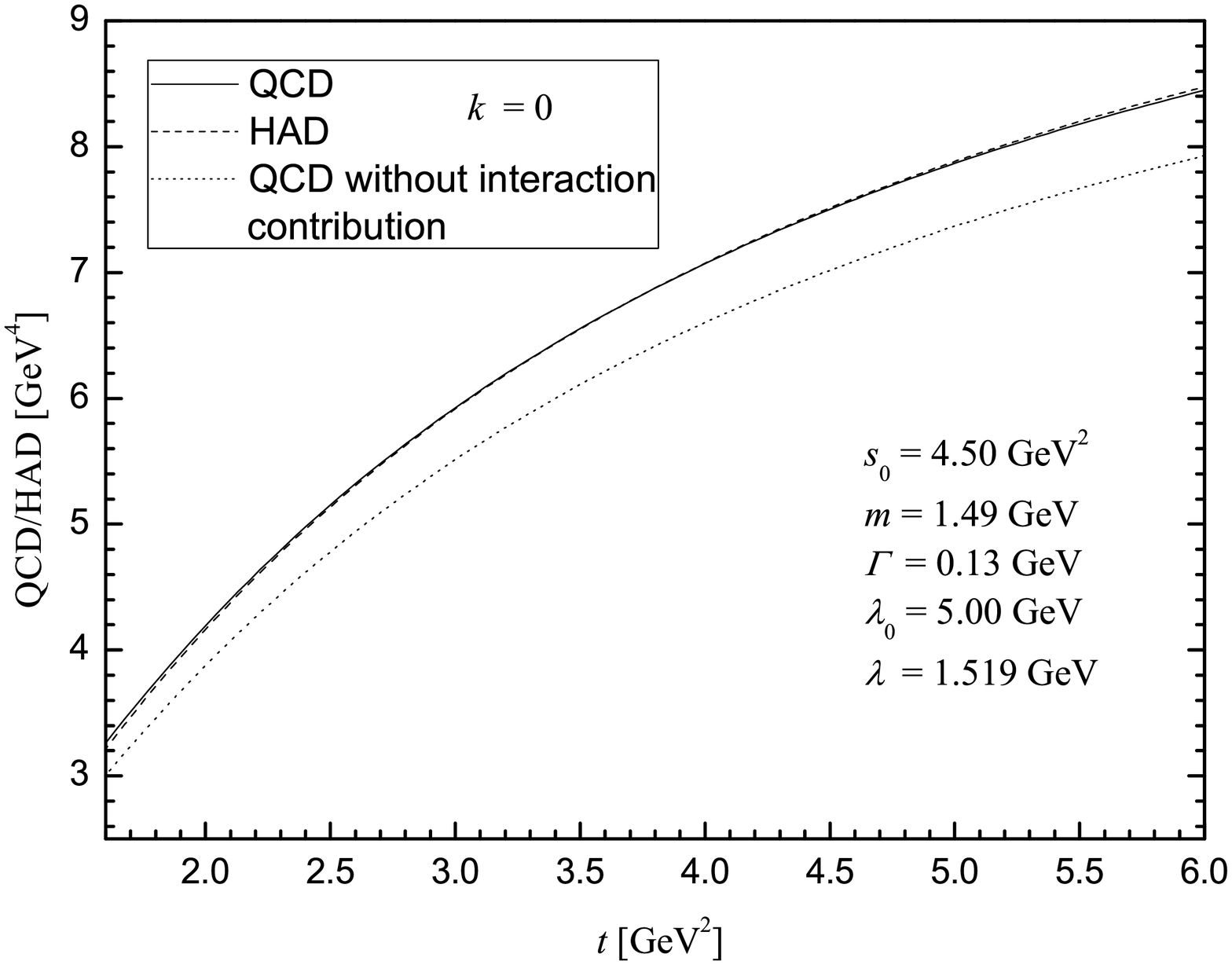}\\
\includegraphics*[angle=0,width=8.5cm]{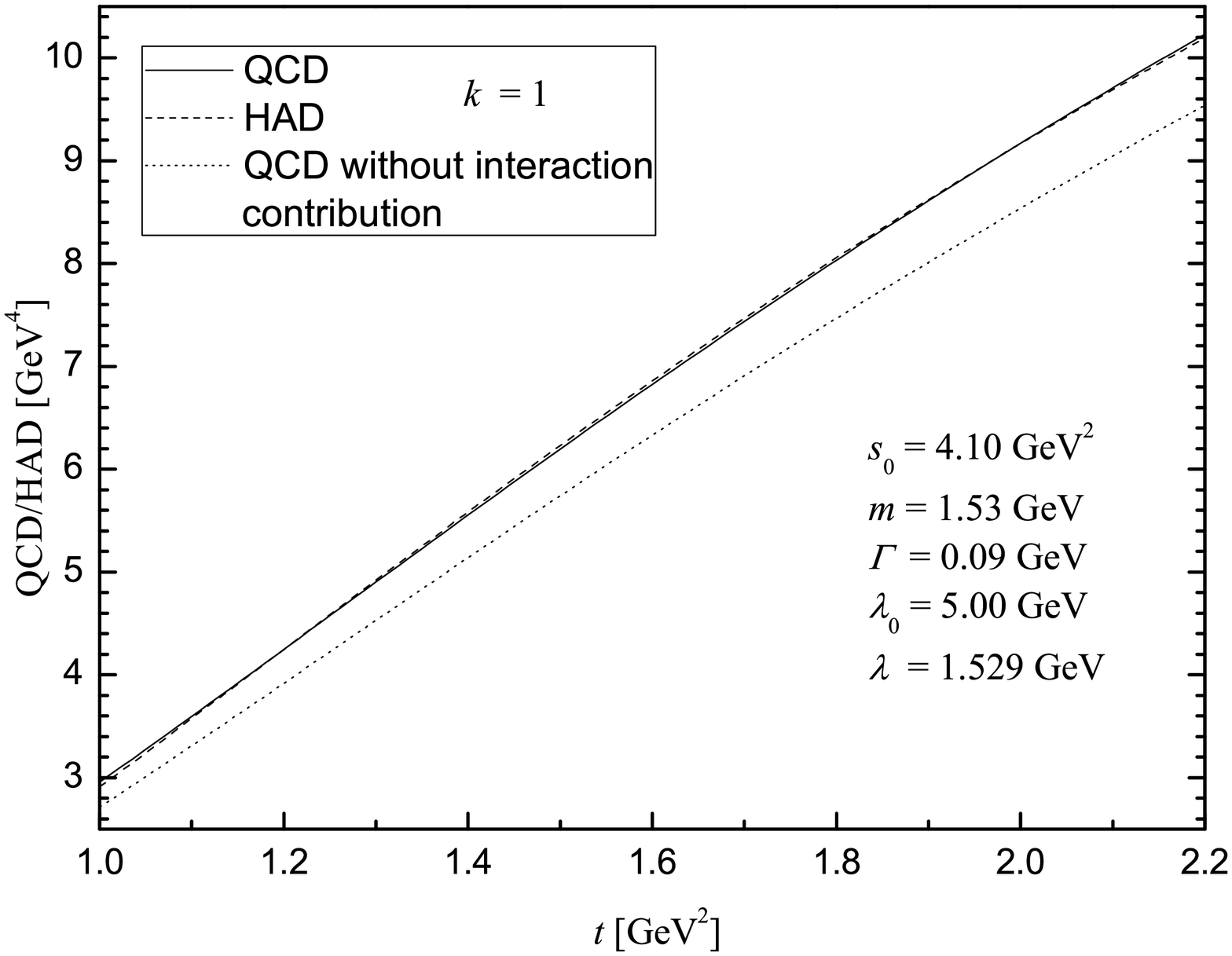}
\caption{\small The curves for the l.h.s. and r.h.s. of the Eq.
(\ref{eq:LSR}) for $k=-1$, $0$ and $+1$ cases of quarkless QCD with
only the lowest resonance considered. The solid line denotes the
r.h.s. (QCD), dashed line for l.h.s. (HAD), and dotted line for the
r.h.s. (QCD) the without interaction contribution.}\label{fig:LQCD}
\end{center}
\end{figure}

\begin{figure}[!hbp]
\begin{center}
\includegraphics*[angle=0,width=8.5cm]{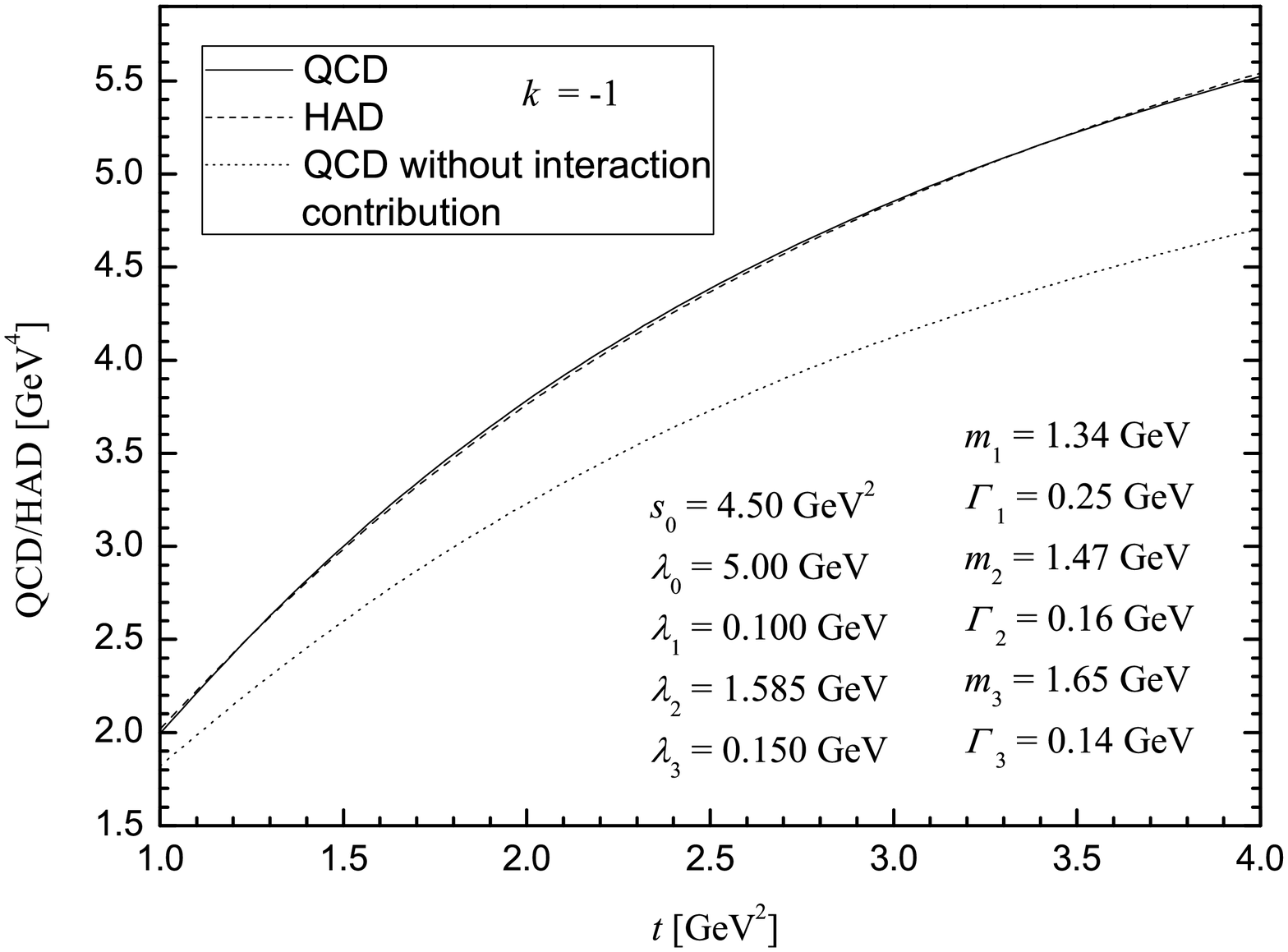}
\includegraphics*[angle=0,width=8.5cm]{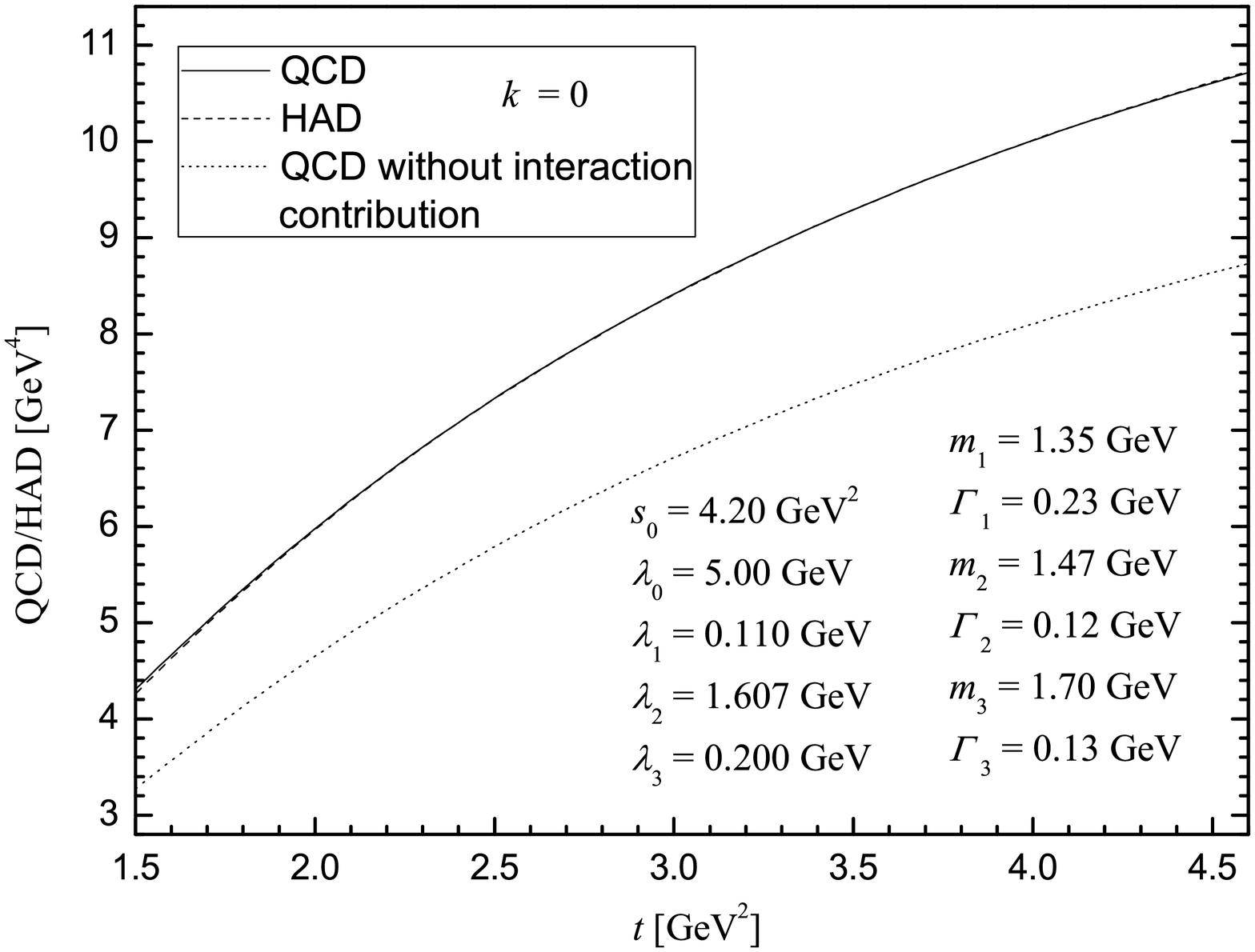}
\includegraphics*[angle=0,width=8.5cm]{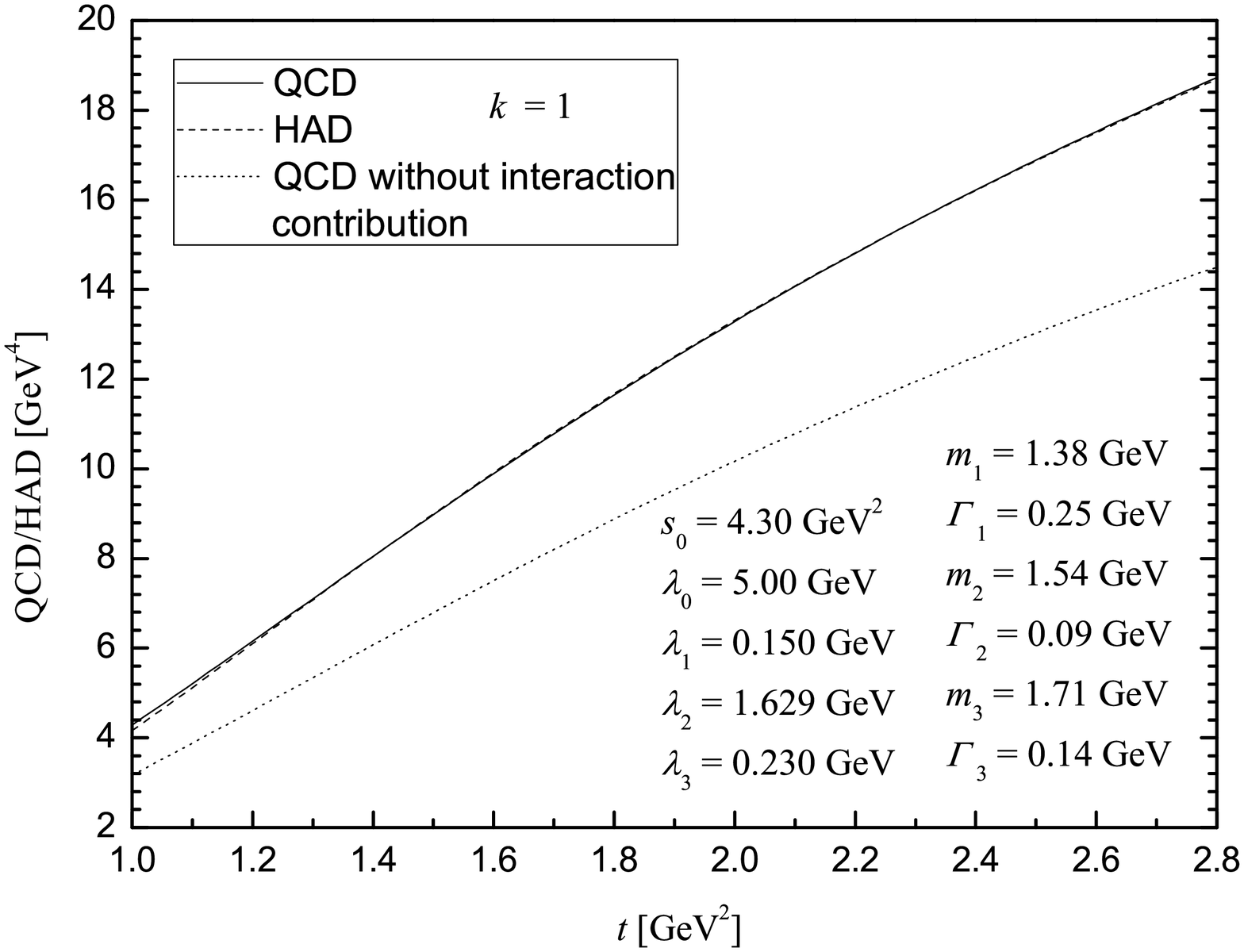}
\caption{\small The curves for the l.h.s. and r.h.s. of the Eq.
(\ref{eq:LSR}) for $k=-1$, $0$ and $+1$ cases of QCD with three
massless quarks and the lowest three resonances considered. The
solid line denotes the r.h.s. (QCD), dashed line for l.h.s. (HAD),
and dotted line for the r.h.s. (QCD) the without interaction
contribution.}\label{fig:LQCD1}
\end{center}
\end{figure}

In the world of quarkless QCD there is only one well defined scalar
bound state of gluons below 1GeV suggested by lattice QCD, and thus
we choose $n=1$ in Eq.(\ref{eq:had2}). Including quarks enhances the
difficulty of the task since many states possessing the same quantum
numbers may present in the correlator. The assumption of a single
well-isolated lowest resonance is questioned from the admixture with
quarkonium states, and from the experimental data that three
$0^{++}$ scalar states around the mass scale of $1500$MeV, namely
$f_0(1370)$, $f_0(1500)$ and $f_0(1710)$. Therefore, we choose $n=3$
for $\rho^{\textrm{had}}(s)$. With the requirements mentioned above,
the optimal parameters governing the sum rules are listed in Tab.\ref{tab:3BWR}.
\begin{table}[!h]
 \caption{\small The fitting values of the mass $m$(GeV) and width $\Gamma$(GeV) of the possible
 $0^{++}$ resonances, and of the parameters $\lambda$(GeV), $f$(GeV) and
 $s_0$(GeV$^2$) characterizing the couplings to the corresponding resonances and the
 continuum threshold, as well as of the sum rule window of $t$(GeV$^2$) and the
 compatibility measure $\delta$ for finite-width Laplace sum rules (\ref{eq:LSR})
 in the quarkless QCD (shown in the first three lines), as well as in QCD with three massless quarks
 (shown in the remainder lines, where the first three lines for $n=1$, and the others for $n=3$).}
\begin{indented}
\item[]\begin{tabular}{cccccccc}
\br
 $k$& $m$& $\Gamma$ & $\lambda$ & $f$
 & $s_0$ & $[t_{\textrm{min}},t_{\textrm{max}}]$ &  $\delta$ ($10^{-5}$)\\
\mr
 $-1$& 1.47& 0.16& 1.525& 1.538& 4.1& $[1.3,3.0]$& $4.39$\\
 0& 1.49& 0.13& 1.519& 1.532& 4.5& $[1.6,6.0]$& $2.32$\\
 1& 1.53& 0.09& 1.529& 1.540& 4.1& $[1.0,2.2]$& $2.23$\\
\mr
 $-1$& 1.47& 0.16& 1.586& 1.598& 4.4& $[1.0,4.0]$& $1.25$\\
 0& 1.48& 0.13& 1.612& 1.624& 4.3& $[1.5,4.6]$& $1.63$\\
 1& 1.52& 0.09& 1.620& 1.632& 4.9& $[1.0,2.5]$& $5.12$\\
\mr
    & 1.34& 0.25& 0.100& 0.452& & \\
 $-1$& 1.47& 0.16& 1.585& 1.597& 4.5& $[1.0,4.0]$& $1.67$\\
    & 1.65& 0.14& 0.150& 0.455& & \\
\mr
    & 1.35& 0.23& 0.110& 0.451& & \\
   0& 1.47& 0.12& 1.607& 1.619& 4.2& $[1.5,4.5]$& $0.59$ \\
    & 1.70& 0.13& 0.200& 0.463& & \\
\mr
    & 1.38& 0.25& 0.150& 0.456& & \\
   1& 1.54& 0.09& 1.629& 1.640& 4.3& $[1.0,2.8]$& $4.69$ \\
    & 1.71& 0.14& 0.230& 0.469& & \\
\br \label{tab:3BWR}
\end{tabular}
\end{indented}
\end{table}
The corresponding curves for the l.h.s. and r.h.s. of (\ref{eq:LSR})
of $k=-1$, $0$ and $+1$ in quarkless QCD for $n=1$, and in QCD with
three massless quarks for $n=3$ are displayed in Figs.\ref{fig:LQCD}
and \ref{fig:LQCD1}, respectively. These figurations show the
consistent match between the both sides of Eq. (\ref{eq:LSR}) for
$k=-1,0$ and $1$ respectively, with the corresponding fitting
parameters. The solid lines are the r.h.s.(QCD) of Eq.
(\ref{eq:LSR}), and the dashed lines are the l.h.s.(HAD) of Eq.
(\ref{eq:LSR}), while the dotted line for the r.h.s. (QCD) excluding
the contribution of interactions between the instantons and the
quantum gluons. The matching between both sides of the sum rules are
very well over the whole fiducial region with a very little
departure. In the case of QCD with three massless quarks, the curves
for $n=1$ are similar to those for $n=3$ with little worse
compatibility, and not displayed here.

\section{Conclusion and discussion}
The properties of $0^{++}$ glueball are examined in a family of the
finite-width Laplace sum rules. The correlation function is
calculated in a semiclassical expansion, a well-defined process
justified in the path-integral quantization formalism, of QCD in the
instanton background, namely the instanton liquid model of the QCD
vacuum. Besides the contributions from pure gluons and instantons
separately, the one arising from the interactions between the
classical instanton fields and the quantum gluon ones are taken into
account as well. Instead of using the usual zero-width approximation
for the spectral function of the considered resonances, the
Breit-Wigner form for the resonances with a correct threshold
behavior is adopted. With the QCD standard input parameters, three
Laplace sum rules with $k=-1,0,1$ are carefully studied.

By taking the average for the values, listed in tab. 1, of the
corresponding $k=-1$, $0$ and $1$ sum rules, for the case of $n=1$,
the values of the mass and width, and the other optical fit
parameters are
\begin{eqnarray}
 &&m=1.51\pm0.15\textrm{GeV},\, \Gamma=0.12\pm 0.04\textrm{GeV}, \nonumber\\
 &&f=1.52\pm0.12\,\textrm{GeV},\,s_0=4.5\pm0.5\,\textrm{GeV}^2
    \nonumber
\end{eqnarray}
in quarkless QCD, and
\begin{eqnarray}
 &&m=1.50\pm0.15\textrm{GeV},\, \Gamma=0.12\pm 0.04\textrm{GeV}, \nonumber\\
 &&f=1.61\pm0.11\,\textrm{GeV},\,s_0=4.5\pm0.5\,\textrm{GeV}^2
    \nonumber
\end{eqnarray}
for QCD with three massless flavors, where the errors are estimated
from the uncertainties of the spread between the individual sum
rules, and by varying the phenomenological parameters, $\Lambda$ and
$\langle\alpha_sG^2\rangle$, appropriately away from their central
values $\Lambda=120-200\,\textrm{MeV}$ and
$\langle\alpha_sG^2\rangle=0.6-0.8\textrm{GeV}^4$. It is remarkable
to notice that these two results are close in value, and it
indicates that the considered quark effects may be not so large at
the energy scale of the resonance mass above 1GeV.

The above conclusion is further justified in the investigation of
the case of $n=3$. We can see that the current is coupled mainly to the resonance state near
$f_0(1500)$ predicted in a single resonance approach. To be
quantitative, let us consider the case with the most excellent
compatibility, namely the $k=0$ results shown in Tab.\ref{tab:3BWR} and Fig.\ref{fig:LQCD1}. The
corresponding couplings $f^6$ to the three resonances $R_1$, $R_2$
and $R_3$ with masses $1.35$GeV, $1.47$GeV and $1.70$GeV are
\begin{equation}
0.092\textrm{GeV}^3,\,\,\,4.073\textrm{GeV}^3,\,\,\,0.094\textrm{GeV}^3
  \label{eq:3couplings}
\end{equation}
respectively, Note that
\begin{equation}
\left(
\begin{array}{l}
\langle0|O_s|R_1\rangle\\
\langle0|O_s|R_2\rangle\\
\langle0|O_s|R_3\rangle
\end{array}
\right)
=M
\left(
\begin{array}{l}
\langle0|O_s|1\rangle\\
\langle0|O_s|8\rangle\\
\langle0|O_s|G\rangle
\end{array}
\right)
\end{equation}
where $M$ stands for the mixing matrix (s. the second
one of Eqs.(45) in Ref.\cite{CM09}). The values of the couplings
of $O_s$ to the $\bar{q}q$ states $|1\rangle$, $|8\rangle$ and the pure glueball
state $|G\rangle$ are
\begin{eqnarray}
&&\langle0|O_s|1\rangle=-0.48\,\textrm{GeV}^3,\nonumber\\
&&\langle0|O_s|8\rangle=0.19\,\textrm{GeV}^3,\\
&&\langle0|O_s|G\rangle=0.86\,\textrm{GeV}^3\nonumber
\end{eqnarray}
after normalization, respectively. Although the estimation
is relatively rough, it is still remarkable to notice
that, first, the coupling to $|G\rangle$ is dominant; and second,
the signs of the couplings to $|1\rangle$ and $|8\rangle$ is consistent
with the scalar glueball-meson coupling theorems\cite{NSVZ80,NSVZ81,Kisslinger08}.

As summary, we may conclude that the values of the mass and decay
width of the $0^{++}$ resonance, in which the fraction of the scalar
glueball state is dominant, are $m=1.5\pm0.15$ GeV and
$\Gamma=0.12\pm 0.03$ GeV, respectively, and the value of its
coupling to the corresponding current is $\lambda_{s\geq
m^2_{\pi}}=1.5\pm0.12\ \,\textrm{GeV}$. They are not only compatible
with lattice QCD simulation
\cite{Colin99,Meng09,Chen06,Vaccar99,Sexton95} and other
estimation\cite{Forkel05, Forkel01, Sch95, Narison98, Liang}, but
also in good accordance with the experimental data of $f_0(1500)$
\cite{Amsler08,Amsler95,Amsler195}.

It is also remarkable that the three Laplace sum rules lead to
almost the same results, a consistency between the subtracted and
unsubtracted sum rules are very well justified. We note that we have
not working within the mixed scheme, namely with including
condensates, and in the same time, adopting the so-called direct
instanton approximation, but simply with a self-consistent
framework, a quantum theory in a classical background, without the
problem of double counting. In this aspect, our results further
justified the instanton liquid model for QCD among other many
justifications.

In our semiclassical expansion, the leading contribution to the sum
rules comes from instantons themselves, especially in the region
below the threshold $s_0$. It is the amount of this contribution
determining the low-bound of the sum rule window. This means that
the non-linear configurations of gluons have a dominant role with
respect to the quantum fluctuations in the low-energy region.

The contribution of the interactions between the classical instanton
fields and quantum gluon ones, considered in this letter but
neglected in earlier sum rule calculations
\cite{Forkel05,Forkel01,HSE01,HaS01,Narison98}, is in fact not
negligible. To the contrast, its amount is approximately double or
even triple of that from the pure quantum fluctuations in the whole
fiducial domain, expected from a view point of the semiclassical
expansion. Moreover, it is obviously seen from Figs.
\ref{fig:LQCD}-\ref{fig:LQCD1} that, without taking the contribution
from the interactions between instantons and quantum gluons into
account, the departures between
$\mathcal{L}^{\textrm{had}}_{k}(s_0,t)$ and
$\mathcal{L}^{\textrm{QCD without interaction}}_{k}(s_0,t)$ become
large, and all the three Laplace sum rules become less stable, and
thus less reliable.

Finally, it should be noticed that the imaginary part of instanton
contribution is an oscillating, amplificatory and nonpositive
defined function, and so is the imaginary part of the correlation
function. This property which is a fatal problem for the QCD sum
rule calculation with the instanton background, may make the
contribution of continuum too large to be under control. Hilmar
Forkel introduced a Gaussian distribution for instanton to get rid
off this trouble, and obtained a smaller $0^{++}$ mass scale:
$1.25\pm0.2$ GeV \cite{Forkel05} compared to the earlier result
$1.53\pm0.2$ GeV \cite{Forkel01}. We didn't use this Gaussian
distribution, but simply choose a smaller fitting parameter to avoid
this problem.

\section*{Acknowledgments}
We are grateful to Prof. H. G. Dosch for useful discussions. This
work is supported by the National Natural Science Foundation of
China under Grant No. 10075036, BEPC National Laboratory Project
R\&D and BES Collaboration Research Foundation,Chinese Academy of
Sciences (CAS) Large-Scale Scientific Facility Program.

\end{document}